\documentclass[aps,preprint,a4paper,showpacs,amsmath,amssymb,floatfix]{revtex4}
\usepackage{graphicx}
\usepackage{dcolumn}
\usepackage{bm}
\newcommand{\Op}[1]{{\boldsymbol{\mathrm{\hat{#1}}}}}
\newcommand{\Opt}[1]{{\boldsymbol{\mathrm{\tilde{#1}}}}}
\newcommand{\beq}{\begin{equation}}
\newcommand{\eeq}{\end{equation}}
\newcommand{\beqar}{\begin{eqnarray}}
\newcommand{\eeqar}{\end{eqnarray}}
\newcommand{\bea}{\begin{eqnarray}}
\newcommand{\eea}{\end{eqnarray}}
\newcommand{\bcen}{\begin{center}}
\newcommand{\ecen}{\end{center}}

\begin{document}

\title{The Quantum Absorption  Refrigerator }

\author{Amikam Levy and Ronnie Kosloff}

\affiliation{
Institute  of Chemistry
The Hebrew University, Jerusalem 91904, Israel\\
}

\begin{abstract}
A quantum absorption refrigerator driven by noise is studied with the purpose
of determining the limitations of cooling to absolute zero. 
The model consists of a working medium coupled simultaneously to hot, cold and noise baths.
Explicit expressions for the cooling power are obtained for Gaussian and Poisson white noise.
The quantum model is consistent with the first and second laws of thermodynamics.
The third law is quantified, the cooling power ${\cal J}_c$ vanishes as ${\cal J}_c \propto T_c^{\alpha}$,
when $T_c \rightarrow 0$, where $\alpha =d+1$ for dissipation by emission and absorption of quanta described by 
a linear coupling to a thermal bosonic field, where $d$ is the dimension of the bath.
\end{abstract}
\maketitle
\section{Introduction} 
\label{sec:introduction}

The adsorption chiller is a refrigerator which employs a heat source to replace mechanical work for
driving a heat pump \cite{jeff00}.  The first device was developed in 1850 by the Carr\'e brothers  which became the first useful refrigerator.
In 1926 Einstein and Szil\'ard invented an absorption refrigerator with no moving parts \cite{szilard1926}. This idea has been incorporated recently to an autonomous quantum absorption refrigerator with no external intervention \cite{k169,popescu10} .
The present study is devoted to a quantum absorption refrigerators driven by noise. The objective is to study the scaling of the optimal
cooling power when the absolute zero temperature is approached.

This study is embedded in the field of {\em Quantum thermodynamics}, the study of thermodynamical processes within the context of quantum dynamics. 
Historically, consistence with  thermodynamics led to Planck's law,  the basics of quantum theory. 
Following the ideas of Planck on black body radiation,  Einstein five years later (1905), quantized the electromagnetic  field \cite{einstein05}.
{\em Quantum thermodynamics} is devoted to unraveling  the intimate connection between the 
laws of thermodynamics and their quantum origin \cite{geusic67,spohn78,alicki79,k24,k122,k156,k169,lloyd,kieu04,segal06,bushev06,erez08,mahler08,allahmahler08,segal09,he09,mahlerbook,popescu10}. In this tradition the present study is aimed toward the quantum study of
the third law of thermodynamics \cite{nerst06,landsberg56}, in particular quantifying  the unattainability principle \cite{belgiorno03}:
What is the scaling of the cooling power ${\cal J}_c$ of a refrigerator when  the cold bath temperature approaches the absolute zero  ${\cal J}_c \propto T_c^{\alpha}$ when $T_c \rightarrow 0$. 
 
\section{The quantum trickle} 
\label{sec:trickle}

\begin{figure}[htbp]
\center{\includegraphics[height=6cm]{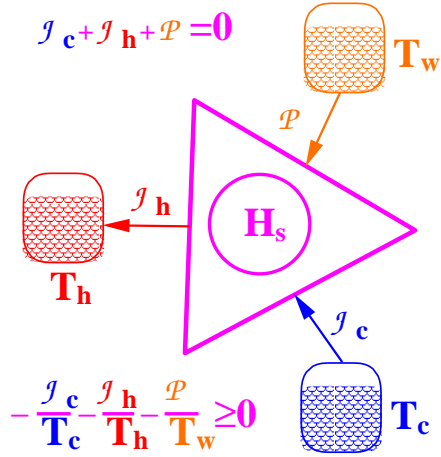}}
\caption{The quantum trickle: A quantum heat pump designated by the Hamiltonian $\Op H_s$
coupled to a work reservoir with temperature $T_w$, a hot reservoir with temperature $T_h$ 
and a cold reservoir with temperature $T_c$. The heat and work currents are indicated. In steady state
${\cal J}_h+{\cal J}_c+{\cal P}=0$.}
\label{fig:1}
\end{figure}

The minimum requirement for a quantum thermodynamical device is a system connected simultaneously 
to three reservoirs \cite{berry84}.
These baths are termed hot, cold and work reservoir  as described in Fig. \ref{fig:1}.
A quantum description requires a representation of the dynamics working medium and the three heat reservoirs.
A reduced description is employed in which the dynamics of the working 
medium is described by the Heisenberg  equation for the operator $\Op O$ for open systems \cite{lindblad76,breuer}:
\begin{equation}
\frac{d}{dt} \Op O ~~=~~ \frac{i}{\hbar} [ \Op H_s, \Op O ] +\frac{\partial \Op O}{\partial t}+ {\cal L}_h (\Op O)+ {\cal L}_c (\Op O)+ {\cal L}_w (\Op O)~,
\label{eq:lvn}
\end{equation}
where $\Op H_s$ is the system Hamiltonian  and ${\cal L}_g$ are the dissipative completely positive superoperators for each bath ($g=h,c,w$).
A minimal  Hamiltonian describing the essence of the quantum refrigerator is composed of three interacting oscillators:
\begin{eqnarray}
\begin{array}{rcl}
\Op H_s &= &\Op H_0 ~+~ \Op H_{int}\\
\Op H_0 &=& \hbar \omega_h \Op a^{\dagger} \Op a +\hbar \omega_c  \Op b^{\dagger}\Op b +\hbar \omega_w \Op c^{\dagger} \Op c \\
\Op H_{int}&=& \hbar \omega_{int} \left( \Op a^{\dagger} \Op b \Op c + \Op a \Op b^{\dagger}  \Op c^{\dagger} \right)~.
\end{array}
\label{eq:hamil}
\end{eqnarray}
$\Op H_{int}$ represents an annihilation of excitations on the work and cold bath simultaneous 
with creating an excitation in the hot bath.
In an open quantum system the superoperators ${\cal L}_g$ 
represent a thermodynamic isothermal partition allowing 
heat flow from the bath to the system. Such a partition is equivalent to the weak coupling limit between the system and bath \cite{k122}. 
The superoperators  ${\cal L}_g$ are derived from the Hamiltonian:
\begin{equation}
\Op H = \Op H_s+\Op H_h+\Op H_c+\Op H_w +\Op H_{sh} + \Op H_{sc}+ \Op H_{sw}~,
\label{eq:hamil1}
\end{equation}
where $\Op H_g$ are bath Hamiltonians and $\Op H_{sg}$ represent system bath coupling.
Each of the oscillators is linearly coupled to a heat reservoir for example for the hot bath: 
$\Op H_{sh} = \lambda_{sh} ( \Op a \Op A_h^{\dagger} + \Op a^{\dagger} \Op A_h)$ . 
Each reservoir individually should equilibrate the working medium
to thermal equilibrium with the reservoir temperature. 
In general, the derivation of a thermodynamically consistent master equation is technically very difficult \cite{alicki06}. 
Typical problems are approximations that violate the laws of thermodynamics.
We therefore require that the master equations fulfill the thermodynamical laws.
Under steady state conditions of operation they become:
\begin{eqnarray}
\begin{array}{rcl}
{\cal J}_h+{\cal J}_c+{\cal P}&=&0\\
-\frac{{\cal J}_h}{T_h}-\frac{{\cal J}_c}{T_c}-\frac{{\cal P}}{T_w} &\ge& 0~,
\end{array}
\label{eq:thermo}
\end{eqnarray}
where ${\cal J}_k = \langle {\cal L}_k (\Op H) \rangle$.
The first equality represents conservation of energy (first law) \cite{spohn78,alicki79}, and the second inequality 
represents positive entropy production in the universe $\Sigma_u \ge 0$ (second law). 
For refrigeration $T_w \ge T_h \ge T_c$. From the second law the scaling exponent $\alpha \ge 1$ \cite{k156}.

\section{Noise driven refrigerator }
{\bf Gaussian noise driven refrigerator}. In the absorption refrigerator the noise source replaces the work bath and its contact $ \hbar \omega_w \Op c^{\dagger} \Op c$ 
leading to:
\begin{eqnarray}
\begin{array}{rcl}
\Op H_{int}&=&  f(t)\left( \Op a^{\dagger} \Op b + \Op a \Op b^{\dagger}   \right)=  f(t) \Op X ~,
\end{array}
\label{eq:hamil2}
\end{eqnarray}
where $f (t) $ is the noise field. $\Op X=(\Op a^{\dagger} \Op b  + \Op a \Op b^{\dagger})$ is
the generator of a swap operation between the two oscillators and  is part of  a set of $SU(2)$ operators ,
$\Op Y=i(\Op a^{\dagger} \Op b  - \Op a \Op b^{\dagger})$,
$\Op Z = \left( \Op a^{\dagger}  \Op a  -   \Op b^{\dagger} \Op b \right)$ and the Casimir
$\Op N = \left( \Op a^{\dagger}  \Op a  +   \Op b^{\dagger} \Op b \right)$.

We first study a Gaussian source of white noise characterized by
zero mean $\langle f(t) \rangle=0$ and delta time correlation $\langle f(t) f(t') \rangle = 2 \eta \delta(t-t')$. The Heisenberg  equation for a time independent operator $\Op O$ reduced to: 
\begin{equation}
\frac{d}{dt} \Op O ~~=~~ i [ \Op H_s, \Op O ]  +{\cal L}_n(\Op O)+{\cal L}_h (\Op O)+ {\cal L}_c (\Op O)~,\label{eq:glvn}
\end{equation}
where $\Op H_s = \hbar \omega_h \Op a^{\dagger} \Op a +\hbar \omega_c  \Op b^{\dagger}\Op b $. 
The noise dissipator for Gaussian noise is ${\cal L}_n(\Op O) = -\eta [ \Op X, [\Op X , \Op O]]$ \cite{gorini76}. 
 
The next step is to derive the quantum Master equation of each reservoir. 
We assume that the reservoirs are uncorrelated and also
uncorrelated with the driving noise. These conditions simplify the derivation of  ${\cal L}_h$ 
which become the standard energy relaxation terms driving oscillator 
$ \omega_h \Op a^{\dagger} \Op a$ to thermal equilibrium with temperature $T_h$ and ${\cal L}_c$ 
drives oscillator $ \hbar \omega_b \Op b^{\dagger} \Op b$ to equilibrium $T_c$ \cite{breuer}.
\begin{eqnarray}
\begin{array}{rcr}
{\cal L}_h (\Op O )&~=~& \Gamma_h (N_h +1 ) \left( \Op a^{\dagger} \Op O  \Op a -\frac{1}{2} \left\{\Op a^{\dagger}   \Op a, \Op O \right\} \right)\\
&&~+~\Gamma_h N_h \left( \Op a \Op O  \Op a^{\dagger}  -\frac{1}{2} \left\{\Op a  \Op a^{\dagger} , \Op O \right\} \right)\\
{\cal L}_c (\Op O )&~=~& \Gamma_c (N_c +1 ) \left( \Op b^{\dagger} \Op O  \Op b -\frac{1}{2} \left\{\Op b^{\dagger}   \Op b, \Op O \right\} \right)\\
&&~+~\Gamma_c N_c \left( \Op b \Op O  \Op b^{\dagger}  -\frac{1}{2} \left\{\Op b  \Op b^{\dagger} , \Op O \right\} \right)\\
\end{array}~.
\label{eq:relaxabsor}
\end{eqnarray}
In the absence of the stochastic driving field these equations drive oscillator $a$ and $b$ separately
to thermal equilibrium  provided that 
$N_h = (\exp(\frac {\hbar \omega_h}{k T_h})-1)^{-1} $ and $N_c = (\exp(\frac {\hbar \omega_c}{k T_c})-1)^{-1} $. The kinetic coefficients $\Gamma_{h/c}$ are 
determined from the baths density function \cite{k122}.

The equations of motion are closed to the $SU(2)$ set of operators. 
To derive the cooling current ${\cal J}_c= \langle {\cal L}_c( \hbar \omega_c \Op b^{\dagger} \Op b)\rangle$,
we solve for stationary solutions of $\Op N$ and $\Op Z$, obtaining: 
\begin{eqnarray}
\begin{array}{rcl}
{\cal J}_c &~=~&  \hbar \omega_c\frac{(N_c-N_h)}{(2\eta)^{-1} +\Gamma_h^{-1} + \Gamma_c^{-1}}
\end{array}~.
\label{eq:Jc}
\end{eqnarray}
Cooling occurs for $N_c > N_h \Rightarrow \frac{\omega_h}{T_h} > \frac{\omega_c}{T_c}$. 
The coefficient of performance ($COP$) for the absorption chiller is defined by the relation 
$COP = \frac{{\cal J}_c}{{\cal J}_n}$, 
with the help of Eq. (\ref{eq:Jc}) we obtain the Otto cycle $COP$ \cite{jahnkemahler08}:
 \begin{equation}
COP ~=~ \frac{\omega_c}{\omega_h - \omega_c} ~\le~ \frac{T_c}{T_h-T_c}~.
\label{eq:COPstoch}
\end{equation}
A different viewpoint starts from the high temperature limit of the work bath $T_w$ based on the weak coupling 
limit in Eq. (\ref{eq:hamil}), (\ref{eq:hamil1}), then:
\begin{eqnarray}
\begin{array}{rcr}
{\cal L}_w (\Op O )&~=~& \Gamma_w (N_w +1 ) \left( \Op a^{\dagger} \Op b \Op O  \Op b^{\dagger} \Op a 
-\frac{1}{2} \left\{\Op a^{\dagger}   \Op a \Op b \Op b^{\dagger} , \Op O \right\} \right)\\
&&~+~\Gamma_w N_w \left( \Op a \Op b^{\dagger} \Op O  \Op a^{\dagger} \Op b 
-\frac{1}{2} \left\{\Op a  \Op a^{\dagger} \Op b^{\dagger} \Op b , \Op O \right\} \right)\\
\end{array}~.
\label{eq:relaxw}
\end{eqnarray}
where $N_w = (\exp(\frac {\hbar \omega_w}{k T_h})-1)^{-1} $. At finite temperature ${\cal L}_w(\Op O)$ does not lead to a close set of equations. But in the limit of $T_w \rightarrow \infty$ it becomes equivalent to the Gaussian noise generator: 
${\cal L}_w (\Op O)= -\eta/2 \left( [ \Op X , [\Op X, \Op O]]+ [ \Op Y , [\Op Y, \Op O]] \right)$, where $\eta= \Gamma_w N_w$.
This noise generator leads to the same current ${\cal J}_c$ and $COP$ as Eq. (\ref{eq:Jc}) and (\ref{eq:COPstoch}).
We conclude that Gaussian noise represents the singular bath limit equivalent to $T_w \rightarrow \infty$. 
As a result the entropy generated by the noise is zero.

The solutions are consistent with the first and second laws of thermodynamics. The $COP$ is restricted by 
the Carnot $COP$. For low temperatures the optimal cooling current can be approximated by ${\cal J}_c \simeq \omega_c \Gamma_c N_c$.
Coupling to a thermal bosonic field such as electromagnetic or acoustic phonons field implies $\Gamma_c \propto \omega_c^{d}$, where $d$ is the heat bath dimension.
Optimizing the cooling current with respect to $\omega_c$ one obtains that the exponent $\alpha$ quantifying the third law ${\cal J}_c \propto T_c^{\alpha}$ is given by $\alpha = d +1$.     

{\bf Poisson noise driven refrigerator}. Poisson white noise can be referred as a sequence of independent random pulses with exponential inter-arrival times. 
These impulses drive the coupling between the oscillators in contact with  the hot and cold bath leading to \cite{luczka91,alicki06}:
\begin{eqnarray}
\label{eq:master-eq}
\begin{array}{rcl}
\dfrac{d \Op O}{dt}&=&(i/\hbar)[\Opt H,\Op O] - (i/\hbar)\lambda \langle \xi \rangle [\Op X,\Op O] \\
 &&+\lambda\left( \int^{\infty}_{-\infty}d\xi P(\xi)e^{(i/\hbar)\xi \Op X}\Op O e^{(-i/\hbar)\xi \Op X} -\Op O \right)~,
 \end{array}
\end{eqnarray}
where $ \Opt H $ is the total Hamiltonian including the baths. $\lambda$ is the rate of events and $\xi$ 
is the impulse strength averaged over a distribution $P(\xi)$.
Using the Hadamard lemma and the fact that the operators form a closed $SU(2)$ algebra, we can separate the noise contribution to its unitary and dissipation parts, leading to the master equation,
\begin{equation}
\label{eq:vn}
\dfrac{d \Op O}{dt}=(i/\hbar)[\Opt H,\Op O]+(i/\hbar)[\Op H^{\prime} ,\Op O]+ {\cal L}_n(\Op O)~.
\end{equation} 
The unitary part is generated with the addition of the Hamiltonian
$ \Op H^{\prime}= \hbar\epsilon \Op X $ with the interaction
\begin{equation}
\epsilon= -\dfrac{\lambda}{2}\int d\xi P(\xi)(2\xi/\hbar -sin(2\xi /\hbar))\nonumber~.
\end{equation}
This term can cause a direct heat leak from the hot to cold bath.
The noise generator ${\cal L}_n(\Op\rho)$,  can be reduced to the form 
$
{\cal L}_n(\Op O )= -\eta [\Op X,[\Op X,\Op O]]~,
$
with a modified noise parameter:
\begin{equation}
\eta=\dfrac{\lambda}{4}\left( 1-\int d\xi P(\xi)cos(2\xi /\hbar)\right) \nonumber~.
\end{equation}
The Poisson noise generates an effective Hamiltonian which is composed of 
$\Opt H$ and $\Op H^{\prime}$, modifying the energy levels of the working medium.
This  new Hamiltonian structure has to be incorporated in the derivation of the master equation 
otherwise the second law will be violated. 
The first step is to rewrite the system Hamiltonian in its dressed form.
A new set of bosonic operators is defined
\begin{eqnarray}
\begin{array}{l}
 \Op A_{1} =  \Op a \cos(\theta) +\Op b \sin(\theta)  \\
 \Op A_{2} =  \Op b \cos(\theta) -\Op a \sin(\theta) ~,
 \end{array}
\end{eqnarray}
The dressed Hamiltonian is given by:
\begin{equation}
\label {eq:dressed H}
 \Op H_{s} = \hbar\Omega_{+}\Op A^{\dagger}_1 \Op A_1 + \hbar\Omega_{-}\Op A^{\dagger}_2 \Op A_2~,
\end{equation}
where
$
\Omega_{\pm} = \dfrac{\omega_h +\omega_c}{2} \pm  \sqrt{(\dfrac{\omega_h -\omega_c}{2})^2 +\epsilon^2} 
$
and
$
\cos^2(\theta) = \dfrac{\omega_h-\Omega_{-}}{\Omega_{+}-\Omega_{-}} 
$
Eq.(\ref {eq:dressed H}) impose the restriction, $\Omega_{\pm}>0$ which can be translated to $\omega_h \omega_c > \epsilon^2$. 
The master equation in the Heisenberg representation becomes:
\begin{equation}
\dfrac{d \Op O}{dt}=(i/\hbar)[\Op H_s ,\Op O] + {\cal L}_h(\Op O) +{\cal L}_c(\Op O)+{\cal L}_n(\Op O)~,
\end{equation}
where 
\begin{eqnarray}
\begin{array}{ll}
{\cal L}_h(\Op O)=&\gamma_1^h \textbf{c}^2(\Op A_1\Op O\Op A_1^{\dagger}-\frac{1}{2}\{\Op A_1\Op A_1^{\dagger},\Op O\})\\
&+\gamma_2^h \textbf{c}^2(\Op A_1^{\dagger}\Op O\Op A_1-\frac{1}{2}\{\Op A_1^{\dagger}\Op A_1,\Op O\})\\ 
&+\gamma_3^h \textbf{s}^2(\Op A_2\Op O\Op A_2^{\dagger}-\frac{1}{2}\{\Op A_2\Op A_2^{\dagger},\Op O\})\\
&+\gamma_4^h \textbf{s}^2(\Op A_2^{\dagger}\Op O\Op A_2-\frac{1}{2}\{\Op A_2^{\dagger}\Op A_2,\Op O\}) \\
{\cal L}_c(\Op O)=&\gamma_1^c \textbf{s}^2(\Op A_1\Op O\Op A_1^{\dagger}-\frac{1}{2}\{\Op A_1\Op A_1^{\dagger},\Op O\})\\
&+\gamma_2^c \textbf{s}^2(\Op A_1^{\dagger}\Op O\Op A_1-\frac{1}{2}\{\Op A_1^{\dagger}\Op A_1,\Op O\})\\ 
&+\gamma_3^c \textbf{c}^2(\Op A_2\Op O\Op A_2^{\dagger}-\frac{1}{2}\{\Op A_2\Op A_2^{\dagger},\Op O\})\\
&+\gamma_4^c \textbf{c}^2(\Op A_2^{\dagger}\Op O\Op A_2-\frac{1}{2}\{\Op A_2^{\dagger}\Op A_2,\Op O\}) 
\end{array}~,
\end{eqnarray}
where $\textbf{s}=\sin(\theta)$ and $\textbf{c}=\cos(\theta)$. And the noise generator:
\begin{equation}
{\cal L}_n(\Op O)= -\eta [\Op W,[\Op W,\Op O]]~,
\end{equation}
where $\Op W=\sin(2 \theta)\Op Z+\cos(2 \theta)\Op X$
and a new set of operators which form an  $SU(2)$ algebra is defined: $\Op X=(\Op 
A_1^{\dagger}\Op A_2+\Op A_2^{\dagger}\Op A_1)$ , $\Op Y=i(\Op A_1^{\dagger}\Op A_2 - \Op A_2^{\dagger}\Op A_1)$  and $\Op Z=(\Op A_1^{\dagger}\Op A_1-\Op A_2^{\dagger}\Op A_2)$. The total number of excitations is accounted for by the operator $\Op N=(\Op A_1^{\dagger}\Op A_1+\Op A_2^{\dagger}\Op A_2)$.  \\ 
The generalized heat transport coefficients become $\zeta_+^k=\gamma_2^k-\gamma_1^k$ and 
$\zeta_-^k=\gamma_4^k-\gamma_3^k$ for $k= h, c$. Applying the Kubo relation \cite{kubo57,kossakowski77}: $\gamma_1^k=e^{-\hbar\Omega_{+}\beta_k}\gamma_2^k$ and $\gamma_3^k=e^{-\hbar\Omega_{-}\beta_k}\gamma_4^k$, leads to the detailed balance relation:
\begin{eqnarray}
\label{eq:occupation}
\begin{array}{l}
\frac{\gamma_1^k}{\zeta_+^k}=\frac{1}{e^{\hbar\Omega_{+}\beta_k}-1}\equiv N_+^k\\
\frac{\gamma_3^k}{\zeta_-^k}=\frac{1}{e^{\hbar\Omega_{-}\beta_k}-1}\equiv N_-^k\nonumber~.
\end{array}
\end{eqnarray}	 

In general $\zeta_{\pm}^k$ is temperature independent and can be calculated specifically for different choices of spectral density of the baths.
For electromagnetic or acoustic phonon field $\zeta_{\pm}^k \propto \Omega_{\pm}^{d}$.
The heat currents ${\cal J}_h$, ${\cal J}_c$ and ${\cal J}_n$ are calculated by solving the equation of motion for the operators at steady state and at the regime of low temperature, where $cos^2(\theta)\approx 1$ and $sin^2(\theta)\approx0$.
\begin{eqnarray}
 \begin{array}{ll}
 \label{eq:equation of motion}
\frac{d \Op N}{dt}=& -\frac{1}{2} (\zeta_+^h +\zeta_-^c) \Op N -\frac{1}{2} (\zeta_+^h -\zeta_-^c) \Op Z +(\zeta_+^h N_+^h +\zeta_-^c N_-^c)  \\ 
\frac{d \Op Z}{dt}=& -\frac{1}{2} (\zeta_+^h +\zeta_-^c) \Op Z -\frac{1}{2} (\zeta_+^h -\zeta_-^c) \Op N +(\zeta_+^h N_+^h -\zeta_-^c N_-^c) -4\eta \Op Z
\end{array} 
\end{eqnarray} 
 
Once the set of linear equations is solved the exact expression for the heat currents is extracted,
${\cal J}_h=\left\langle {\cal L}_h(\Op H_{s})\right\rangle $, $
{\cal J}_c=\left\langle {\cal L}_c(\Op H_{s})\right\rangle $ and $
{\cal J}_n=\left\langle {\cal L}_n(\Op H_{s})\right\rangle $. For simplicity, 
the distribution of impulses in Eq. (\ref{eq:master-eq}),  is chosen as $P(\xi)=\delta (\xi-\xi_0)$. Then
the effective noise parameter becomes:
\begin{equation}
\eta=\frac{\lambda}{4}(1-cos(2\xi_0/\hbar))~. 
\label{eq:eta}
\end{equation}
The energy shift is controlled by:
\begin{equation}
\epsilon=-\frac{\lambda}{2}(2\xi_0 /\hbar-sin(2\xi_0/\hbar))~.
\label{eq:eps}
\end{equation}
 
\begin{figure}[htbp]
\center{\includegraphics[height=6cm]{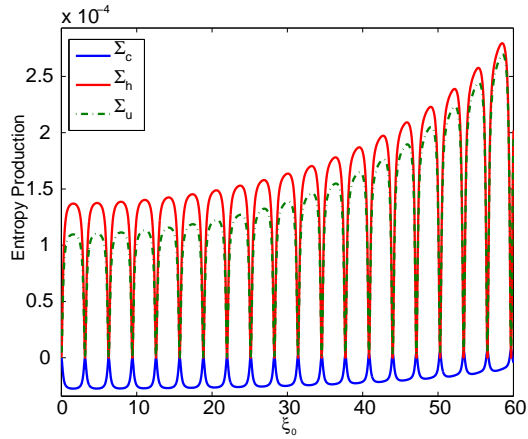}}
\caption{Entropy production $\Sigma_k=-{\cal J}_k/T_k$ as a function of impulse $\xi_0$ for the cold $\Sigma_c$ hot $\Sigma_h$ and the total entropy production $\Sigma_u=\Sigma_h+\Sigma_c$. $ T_c=10^{-3}$, $T_h=2.$, $ \omega_c=T_c$, $\omega_h=10.$ $\lambda=\omega_c$ $\zeta_{\pm}^k=\omega_c/10$ ($\hbar=k=1$).}
\label{fig:2}
\end{figure}	
Figure \ref{fig:2} shows a periodic structure of the heat current ${\cal J}_c$ and  the entropy production 
$\Sigma_c=-{\cal J}_c/T_c$ with the impulse $\xi_0$. The 
second law of thermodynamics is obtained by the balance of the large entropy generation on the hot bath compensating
for the negative entropy generation of cooling the cold bath.
The $COP$ for the Poisson driven refrigerator is restricted by the Otto and Carnot $COP$:
\begin{equation}
COP =\frac{\Omega_-}{\Omega_+ - \Omega_-} \le \frac{\omega_c}{\omega_h-\omega_c} \le \frac{T_c}{T_h-T_c}~.
\label{eq:cop2}
\end{equation}

The heat current ${\cal J}_c$ is given by:
\begin{equation}
{\cal J}_c \approx \hbar \Omega_- \dfrac{N_-^c -N_+^h}{(2\eta)^{-1} + (\zeta_+^h)^{-1} +(\zeta_-^c)^{-1}}~,
\label{eq:pjc}
\end{equation}

The scaling of the optimal cooling rate is now accounted for.  
The heat flow is maximized with respect to the impulse $\xi_0$ by maximizing $\eta$ Eq. (\ref{eq:eta}),
which occurs for $\xi_0= n\frac{\pi}{2}$, ($n=1,2..$). On the other hand  the energy shift $\epsilon^2$ 
Eq. (\ref{eq:eps}) should to be minimized. The optimum is obtained when $\xi_0=\frac{\pi}{2}$. 
The cooling power of the Poisson noise case Eq. (\ref{eq:pjc}) is similar to the Gaussian one Eq. (\ref{eq:Jc}). In the Poisson case also the noise driving parameter $\eta $ is restricted by $ \omega_c$. This is because $\epsilon$ is restricted by
$\Omega_- \ge 0$ and therefore $\lambda$ is restricted to scale with $\omega_c$. 
In total when $T_c \rightarrow 0$,  ${\cal J}_c \propto T_c^{d+1}$.

The optimal scaling relation ${ \cal J}_c \propto T_c^{\alpha}$ of the autonomous absorption refrigerators
should be compared to the scaling of   the discrete  four stroke Otto refrigerators \cite{k243}.
In the driven discrete case the scaling  
depends on the external control scheduling function on the expansion stroke. 
For a scheduling function determined by  a constant frictionless nonadiabatic parameter the 
optimal cooling rate scaled with $\alpha=2$. Faster 
frictionless scheduling procedures were found based on a bang-bang type optimal control solutions.
These solutions led to a scaling of $\alpha=3/2$ when positive frequencies were employed and
${\cal J}_c \propto -T_c/\log T_c$ when negative imaginary frequencies were allowed \cite{muga09,karl11}. 
The drawback of the externally driven refrigerators is that their analysis is complex.  
The optimal scaling assumes that the heat conductivity $\Gamma \gg \omega_c$, 
and that noise in the controls does not influence the scaling. For this reason an analysis based on the
autonomous refrigerators is superior.  
\bibliography{../../Tova/Paper21/dephc1,../../Database/pub}

\begin{thebibliography}{36}
\expandafter\ifx\csname natexlab\endcsname\relax\def\natexlab#1{#1}\fi
\expandafter\ifx\csname bibnamefont\endcsname\relax
  \def\bibnamefont#1{#1}\fi
\expandafter\ifx\csname bibfnamefont\endcsname\relax
  \def\bibfnamefont#1{#1}\fi
\expandafter\ifx\csname citenamefont\endcsname\relax
  \def\citenamefont#1{#1}\fi
\expandafter\ifx\csname url\endcsname\relax
  \def\url#1{\texttt{#1}}\fi
\expandafter\ifx\csname urlprefix\endcsname\relax\def\urlprefix{URL }\fi
\providecommand{\bibinfo}[2]{#2}
\providecommand{\eprint}[2][]{\url{#2}}

\bibitem[{\citenamefont{{J. M. Gordon and K. C. Ng}}(2000)}]{jeff00}
\bibinfo{author}{\bibnamefont{{J. M. Gordon and K. C. Ng}}},
  \emph{\bibinfo{title}{Cool Thermodynamics}} (\bibinfo{publisher}{Cambridge
  International Science Publishing}, \bibinfo{year}{2000}).

\bibitem[{\citenamefont{{A. Einstein and L. Szil\'ard}}(1930)}]{szilard1926}
\bibinfo{author}{\bibnamefont{{A. Einstein and L. Szil\'ard}}},
  \bibinfo{journal}{"US patent No 1,781,541"}  (\bibinfo{year}{1930}).

\bibitem[{\citenamefont{{Jos\'e P. Palao, Ronnie Kosloff, and Jeffrey M.
  Gordon}}(2001)}]{k169}
\bibinfo{author}{\bibnamefont{{Jos\'e P. Palao, Ronnie Kosloff, and Jeffrey M.
  Gordon}}}, \bibinfo{journal}{Phys. Rev. E} \textbf{\bibinfo{volume}{64}},
  \bibinfo{pages}{056130} (\bibinfo{year}{2001}).

\bibitem[{\citenamefont{{N. Linden, S. Popescu, P.
  Skrzypczyk}}(2010)}]{popescu10}
\bibinfo{author}{\bibnamefont{{N. Linden, S. Popescu, P. Skrzypczyk}}},
  \bibinfo{journal}{Phys. Rev. Lett.} \textbf{\bibinfo{volume}{105}},
  \bibinfo{pages}{130401} (\bibinfo{year}{2010}).

\bibitem[{\citenamefont{{A. Einstein}}(1905)}]{einstein05}
\bibinfo{author}{\bibnamefont{{A. Einstein}}}, \bibinfo{journal}{Annalen der
  Physik} \textbf{\bibinfo{volume}{17}}, \bibinfo{pages}{132}
  (\bibinfo{year}{1905}).

\bibitem[{\citenamefont{Geusic et~al.}(1967)\citenamefont{Geusic, du~Bois,
  Grasse, and Scovil}}]{geusic67}
\bibinfo{author}{\bibfnamefont{J.}~\bibnamefont{Geusic}},
  \bibinfo{author}{\bibfnamefont{E.~S.} \bibnamefont{du~Bois}},
  \bibinfo{author}{\bibfnamefont{R.~D.} \bibnamefont{Grasse}},
  \bibnamefont{and} \bibinfo{author}{\bibfnamefont{H.}~\bibnamefont{Scovil}},
  \bibinfo{journal}{Phys. Rev.} \textbf{\bibinfo{volume}{156}},
  \bibinfo{pages}{343} (\bibinfo{year}{1967}).

\bibitem[{\citenamefont{Spohn and Lebowitz}(1978)}]{spohn78}
\bibinfo{author}{\bibfnamefont{H.}~\bibnamefont{Spohn}} \bibnamefont{and}
  \bibinfo{author}{\bibfnamefont{J.}~\bibnamefont{Lebowitz}},
  \bibinfo{journal}{Adv. Chem. Phys.} \textbf{\bibinfo{volume}{109}},
  \bibinfo{pages}{38} (\bibinfo{year}{1978}).

\bibitem[{\citenamefont{Alicki}(1979)}]{alicki79}
\bibinfo{author}{\bibfnamefont{R.}~\bibnamefont{Alicki}}, \bibinfo{journal}{J.
  Phys A: Math.Gen.} \textbf{\bibinfo{volume}{12}}, \bibinfo{pages}{L103}
  (\bibinfo{year}{1979}).

\bibitem[{\citenamefont{{R. Kosloff}}(1984)}]{k24}
\bibinfo{author}{\bibnamefont{{R. Kosloff}}}, \bibinfo{journal}{J. Chem. Phys.}
  \textbf{\bibinfo{volume}{80}}, \bibinfo{pages}{1625} (\bibinfo{year}{1984}).

\bibitem[{\citenamefont{{Eitan Geva and Ronnie Kosloff}}(1996)}]{k122}
\bibinfo{author}{\bibnamefont{{Eitan Geva and Ronnie Kosloff}}},
  \bibinfo{journal}{J. Chem. Phys.} \textbf{\bibinfo{volume}{104}},
  \bibinfo{pages}{7681} (\bibinfo{year}{1996}).

\bibitem[{\citenamefont{{Ronnie Kosloff, Eitan Geva and Jeffrey M.
  Gordon}}(2000)}]{k156}
\bibinfo{author}{\bibnamefont{{Ronnie Kosloff, Eitan Geva and Jeffrey M.
  Gordon}}}, \bibinfo{journal}{J. Appl. Phys.} \textbf{\bibinfo{volume}{87}},
  \bibinfo{pages}{8093} (\bibinfo{year}{2000}).

\bibitem[{\citenamefont{Lloyd}(1997)}]{lloyd}
\bibinfo{author}{\bibfnamefont{S.}~\bibnamefont{Lloyd}},
  \bibinfo{journal}{Phys. Rev. A} \textbf{\bibinfo{volume}{56}},
  \bibinfo{pages}{3374} (\bibinfo{year}{1997}).

\bibitem[{\citenamefont{{T. D. Kieu}}(2004)}]{kieu04}
\bibinfo{author}{\bibnamefont{{T. D. Kieu}}}, \bibinfo{journal}{Phys. Rev.
  Lett.} \textbf{\bibinfo{volume}{93}}, \bibinfo{pages}{140403}
  (\bibinfo{year}{2004}).

\bibitem[{\citenamefont{{D. Segal and A. Nitzan}}(2006)}]{segal06}
\bibinfo{author}{\bibnamefont{{D. Segal and A. Nitzan}}},
  \bibinfo{journal}{Phys. Rev. E} \textbf{\bibinfo{volume}{73}},
  \bibinfo{pages}{026109} (\bibinfo{year}{2006}).

\bibitem[{\citenamefont{{P. Bushev,D. Rotter, A. Wilson, F. Dubin, C. Becher,
  J. Eschner, R. Blatt, V. Steixner, P. Rabl and P. Zoller }}(2006)}]{bushev06}
\bibinfo{author}{\bibnamefont{{P. Bushev,D. Rotter, A. Wilson, F. Dubin, C.
  Becher, J. Eschner, R. Blatt, V. Steixner, P. Rabl and P. Zoller }}},
  \bibinfo{journal}{Phys. Rev. Lett.} \textbf{\bibinfo{volume}{96}},
  \bibinfo{pages}{043003} (\bibinfo{year}{2006}).

\bibitem[{\citenamefont{{E. Boukobza and D. J. Tannor}}(2008)}]{erez08}
\bibinfo{author}{\bibnamefont{{E. Boukobza and D. J. Tannor}}},
  \bibinfo{journal}{Phys. Rev. A} \textbf{\bibinfo{volume}{78}},
  \bibinfo{pages}{013825} (\bibinfo{year}{2008}).

\bibitem[{\citenamefont{{J. Birjukov, T. Jahnke, G. Mahler}}(2008)}]{mahler08}
\bibinfo{author}{\bibnamefont{{J. Birjukov, T. Jahnke, G. Mahler}}},
  \bibinfo{journal}{Eur. Phys. J. B} \textbf{\bibinfo{volume}{64}},
  \bibinfo{pages}{105} (\bibinfo{year}{2008}).

\bibitem[{\citenamefont{{A. E. Allahverdyan, R.S. Johal and G.
  Mahler}}(2008)}]{allahmahler08}
\bibinfo{author}{\bibnamefont{{A. E. Allahverdyan, R.S. Johal and G. Mahler}}},
  \bibinfo{journal}{Phys. Rev. E} \textbf{\bibinfo{volume}{77}},
  \bibinfo{pages}{041118} (\bibinfo{year}{2008}).

\bibitem[{\citenamefont{{D. Segal}}(2009)}]{segal09}
\bibinfo{author}{\bibnamefont{{D. Segal}}}, \bibinfo{journal}{J. Chem. Phys.}
  \textbf{\bibinfo{volume}{130}}, \bibinfo{pages}{134510}
  (\bibinfo{year}{2009}).

\bibitem[{\citenamefont{{H. Wang, SQ Liu , JZ He}}(2009)}]{he09}
\bibinfo{author}{\bibnamefont{{H. Wang, SQ Liu , JZ He}}},
  \bibinfo{journal}{Phys. Rev. E} \textbf{\bibinfo{volume}{79}},
  \bibinfo{pages}{041113} (\bibinfo{year}{2009}).

\bibitem[{\citenamefont{{J. Gemmer, M. Michel and G.
  Mahler}}(2009)}]{mahlerbook}
\bibinfo{author}{\bibnamefont{{J. Gemmer, M. Michel and G. Mahler}}},
  \emph{\bibinfo{title}{{Quantum Thermodynamics}}}
  (\bibinfo{publisher}{Springer}, \bibinfo{year}{2009}).

\bibitem[{\citenamefont{{W. Nernst}}(1906)}]{nerst06}
\bibinfo{author}{\bibnamefont{{W. Nernst}}}, \bibinfo{journal}{{Nachr. Kgl.
  Ges. Wiss. G\"ott.}} \textbf{\bibinfo{volume}{1}}, \bibinfo{pages}{40}
  (\bibinfo{year}{1906}).

\bibitem[{\citenamefont{{P. T. Landsberg}}(1956)}]{landsberg56}
\bibinfo{author}{\bibnamefont{{P. T. Landsberg}}}, \bibinfo{journal}{Rev. Mod.
  Phys.} \textbf{\bibinfo{volume}{28}}, \bibinfo{pages}{363}
  (\bibinfo{year}{1956}).

\bibitem[{\citenamefont{{F. Belgiorno}}(2003)}]{belgiorno03}
\bibinfo{author}{\bibnamefont{{F. Belgiorno}}}, \bibinfo{journal}{J. Phys A:
  Math.Gen.} \textbf{\bibinfo{volume}{36}}, \bibinfo{pages}{8165}
  (\bibinfo{year}{2003}).

\bibitem[{\citenamefont{{Bjarne Andresen, Peter Salamon, and R. Stephen
  Berry}}(1984)}]{berry84}
\bibinfo{author}{\bibnamefont{{Bjarne Andresen, Peter Salamon, and R. Stephen
  Berry}}}, \bibinfo{journal}{"Physics Today"} \textbf{\bibinfo{volume}{37:9}},
  \bibinfo{pages}{{62}} (\bibinfo{year}{1984}).

\bibitem[{\citenamefont{Lindblad}(1976)}]{lindblad76}
\bibinfo{author}{\bibfnamefont{G.}~\bibnamefont{Lindblad}},
  \bibinfo{journal}{Comm. Math. Phys.} \textbf{\bibinfo{volume}{48}},
  \bibinfo{pages}{119} (\bibinfo{year}{1976}).

\bibitem[{\citenamefont{{H.-P. Breuer and F. Petruccione}}(2002)}]{breuer}
\bibinfo{author}{\bibnamefont{{H.-P. Breuer and F. Petruccione}}},
  \emph{\bibinfo{title}{Open quantum systems}} (\bibinfo{publisher}{Oxford
  university press}, \bibinfo{year}{2002}).

\bibitem[{\citenamefont{R.~Alicki and Zanardi}(2006)}]{alicki06}
\bibinfo{author}{\bibfnamefont{D.~L.} \bibnamefont{R.~Alicki}}
  \bibnamefont{and} \bibinfo{author}{\bibfnamefont{P.}~\bibnamefont{Zanardi}},
  \bibinfo{journal}{Phys. Rev. A} \textbf{\bibinfo{volume}{73}},
  \bibinfo{pages}{052311} (\bibinfo{year}{2006}).

\bibitem[{\citenamefont{Gorini and Kossakowski}(1976)}]{gorini76}
\bibinfo{author}{\bibfnamefont{V.}~\bibnamefont{Gorini}} \bibnamefont{and}
  \bibinfo{author}{\bibfnamefont{A.}~\bibnamefont{Kossakowski}},
  \bibinfo{journal}{J. Math. Phys.} \textbf{\bibinfo{volume}{17}},
  \bibinfo{pages}{1298} (\bibinfo{year}{1976}).

\bibitem[{\citenamefont{{T. Jahnke, J. Birjukov and G.
  Mahler}}(2008)}]{jahnkemahler08}
\bibinfo{author}{\bibnamefont{{T. Jahnke, J. Birjukov and G. Mahler}}},
  \bibinfo{journal}{Ann.Phys.} \textbf{\bibinfo{volume}{17}},
  \bibinfo{pages}{88} (\bibinfo{year}{2008}).

\bibitem[{\citenamefont{{J. \'Luczka and M. Niemeic}}(1991)}]{luczka91}
\bibinfo{author}{\bibnamefont{{J. \'Luczka and M. Niemeic}}},
  \bibinfo{journal}{J. Phys A: Math.Gen.} \textbf{\bibinfo{volume}{24}},
  \bibinfo{pages}{L1021} (\bibinfo{year}{1991}).

\bibitem[{\citenamefont{{R. Kubo}}(1957)}]{kubo57}
\bibinfo{author}{\bibnamefont{{R. Kubo}}}, \bibinfo{journal}{J. Phys. Soc.
  Jpn.} \textbf{\bibinfo{volume}{12}}, \bibinfo{pages}{550}
  (\bibinfo{year}{1957}).

\bibitem[{\citenamefont{Kossakowski et~al.}(1977)\citenamefont{Kossakowski,
  Frigerio, Gorini, and Verri}}]{kossakowski77}
\bibinfo{author}{\bibfnamefont{A.}~\bibnamefont{Kossakowski}},
  \bibinfo{author}{\bibfnamefont{A.}~\bibnamefont{Frigerio}},
  \bibinfo{author}{\bibfnamefont{V.}~\bibnamefont{Gorini}}, \bibnamefont{and}
  \bibinfo{author}{\bibfnamefont{M.}~\bibnamefont{Verri}},
  \bibinfo{journal}{Commun. Math. Phys.} \textbf{\bibinfo{volume}{57}},
  \bibinfo{pages}{97} (\bibinfo{year}{1977}).

\bibitem[{\citenamefont{{Yair Rezek, Peter Salamon, Karl Heinz Hoffmann and
  Ronnie Kosloff}}(2009)}]{k243}
\bibinfo{author}{\bibnamefont{{Yair Rezek, Peter Salamon, Karl Heinz Hoffmann
  and Ronnie Kosloff}}}, \bibinfo{journal}{Euro. Phys. Lett.}
  \textbf{\bibinfo{volume}{85}}, \bibinfo{pages}{30008} (\bibinfo{year}{2009}).

\bibitem[{\citenamefont{{Xi Chen, A. Ruschhaupt, S. Schmidt, A. del Campo, D.
  Guery-Odelin, J. G. Muga}}(2010)}]{muga09}
\bibinfo{author}{\bibnamefont{{Xi Chen, A. Ruschhaupt, S. Schmidt, A. del
  Campo, D. Guery-Odelin, J. G. Muga}}}, \bibinfo{journal}{Phys. Rev. Lett.}
  \textbf{\bibinfo{volume}{104}}, \bibinfo{pages}{063002}
  (\bibinfo{year}{2010}).

\bibitem[{\citenamefont{{Karl Heinz Hoffmann, Peter Salamon, Yair Rezek and
  Ronnie Kosloff}}(2011)}]{karl11}
\bibinfo{author}{\bibnamefont{{Karl Heinz Hoffmann, Peter Salamon, Yair Rezek
  and Ronnie Kosloff}}}, \bibinfo{journal}{Eur. Phys. Lett.}
  (\bibinfo{year}{2011}).

\end{thebibliography}

\end{document}